\documentclass[twocolumn,aps,prl,showpacs]{revtex4}
\usepackage{graphicx}
\usepackage{latexsym}
\usepackage{amssymb}
\usepackage{amsmath}
\def\beq{\begin{equation}}
\def\eeq{\end{equation}}
\def\bsy{\boldsymbol}

\def\etal{{\it et al\/.}}

\def\l{{\lambda}}

\def\ket#1{|#1\rangle}
\def\bra#1{\langle#1|}
\def\braket#1#2{\langle#1|#2\rangle}

\begin{document}


\title{Topologically biased random walk with application for community finding in networks}

\author{Vinko Zlati\'{c}$^{1,2}$\thanks{vzlatic@irb.hr}, Andrea 
Gabrielli$^{1,3}$, Guido Caldarelli$^{1,4,5}$}
\affiliation{$^1$Istituto Sistemi Complessi - CNR, UOS ``Sapienza'',
Dipartimento di Fisica, Universit\`a ``Sapienza'', Piazzale A. Moro 2,
00185-Rome, Italy}
\affiliation{$^2$Theoretical Physics Division, Rudjer Bo\v{s}kovi\'{c} 
Institute, P.O.Box 180, HR-10002 Zagreb, Croatia}
\affiliation{$^3$Istituto Sistemi Complessi - CNR, Via dei Taurini 19,
00185-Rome, 
Italy}
\affiliation{$^4$LINKALAB, Via San Benedetto 88, 09129 Cagliari, Italy }
\affiliation{$^5$London Institute for Mathematical Sciences 22 South Audley 
St Mayfair London W1K 2NY, UK}

\begin{abstract}
We present a new approach of topology biased random walks for
undirected networks. We focus on a one parameter family of biases and
by using a formal analogy with perturbation theory in quantum mechanics
we investigate the features of biased random walks.  This analogy is
extended through the use of parametric equations of motion (PEM) to
study the features of random walks {\em vs.} parameter values. Furthermore, we
show an analysis of the spectral gap maximum associated to the value of 
the second eigenvalue of the transition matrix related to the 
relaxation rate to the stationary state. 
Applications of these studies allow {\em ad hoc} algorithms for the exploration 
of complex networks and their communities.
\end{abstract}


\maketitle

\section{Introduction}
The study of complex networks has notably increased in the last years
with applications to a variety of fields ranging from computer
science\cite{broder} and biology to social
science\cite{wiki1,wiki2,catanza} and finance\cite{battiston}. A
central problem in network science \cite{Lovasz93,Aldous} is the study
of the Random Walks (RW) on a graph, and in particular of the relation
between the topological properties of the network and the properties
of diffusion on it.  This subject is not only interesting from a purely
theoretical perspective, but it has important implications to various
scientific issues ranging from epidemics \cite{VPS} to the
classification of web pages through PageRank algorithm \cite{Pr}. Finally, 
RW theory is also used in algorithms for community
detection~\cite{Blanchard, Fortunato, Capocci, alb1, ng1}.

In this paper we set up a new framework for the study of topologically
biased random walks on graphs.  This allows to address problems of
community detection and synchronization~\cite{alb2} in the field of
complex networks \cite{barabasi1,buchanan}.  In particular by using
topological properties of the network to bias the RWs we explore the
network structure more efficiently. A similar approach but with
different focus can be found in \cite{Latora}. In this research we are
motivated by the idea that biased random walks can be efficiently used
for community finding. To this aim we introduce a set of mathematical
tools which allow us an efficient investigation of the ``bias
parameters'' space. We apply this tools to uncover some details in the
spectra of graph transition matrix, and use the relation between
spectra and communities in order to introduce a novel methodology for an
efficient community finding.  The paper is organized as follows: in
the first section we define the topologically biased random walks
(TBRW). We then develop the mathematical formalism used in this paper,
specifically the perturbation methods and the parametric equations of
motion, to track the behaviour of different biases. In the second
section we focus on the behavior of spectral gap in biased random
walks. We define the conditions for which such a spectral gap is
maximal and we present numerical evidence that this maximum is
global. In the third section we present an invariant quantity for the
biased random walk; such constant quantity depends only upon topology
for a broad class of biased random walks.  Finally, in the fourth
section we present a general methodology for the application of different
TBRW in the community finding problems.  We then conclude by providing
a short discussion of the material presented and by providing an
outlook on different possible applications of TBRW.

\section{Biased random walks}

RWs on graphs are a sub-class of Markov chains~\cite{Feller}.  The
traditional approach deals with the connection of the {\em unbiased}
RW properties to the spectral features of \emph{transition operators}
associated to the network \cite{FanChung}.  A generic graph can be represented
by means of the adjacency matrix $\bsy{\hat{A}}$ whose entries
$A_{ij}$ are $1$ if an edge connects vertices $i$ and $j$ and $0$
otherwise.  Here we consider undirected graphs so that $\bsy{\hat{A}}$
is symmetric. 
The \emph{normal matrix} $\bsy{\hat{T}}$ is related to $\bsy{\hat{A}}$
through 
$\bsy{\hat{T}}=\bsy{\hat{A}}\bsy{\hat{k}}^{-1}$,   
where $\bsy{\hat{k}}$ is a diagonal matrix with
$(\bsy{\hat{k}})_{ii}=k_i$, i.e. the
degree, or number of edges, of vertex $i$.  In the following we
use uppercase letters for non-diagonal matrices and lowercase
letters for the diagonal ones.  Note that by definition $k_j=\sum_i A_{ij}$.
Consequently $\sum_i T_{ij}=1$ with $T_{ij}\ne 0$ {\it iif} $A_{ij}=1$, i.e. if
$i$ and $j$ are nearest neighbors vertices. The matrix $\{T_{ij}\}$ defines the
transition probabilities for an
{\em unbiased} random walker to pass from $j$ to $i$. 
In such a case $T_{ij}$ has the same positive value for
any of the neighbors $i$ of $j$ and vanishes for all the other vertices\cite{Noh}.
In analogy to the operator defining the single step transition
probabilities in general Markov chains, $\bsy{\hat{T}}$ is
also called the transition {\em matrix} of the unbiased RW.

A {\em biased} RW on a graph can be defined by a more general transition matrix 
$\hat{\bf T}$ where the element $T_{ij}$ gives again
the probability that a walker on the vertex $j$ of the graph will move
to the vertex $i$ in a single step, but depending on appropriate weights for
each pair of vertex $(i,j)$. 
A genuine way to write these
probabilities is to assign weights $W_{ij}$ which represent the rates of jumps
from vertex $j$ to vertex $i$ and normalize them: 
\beq\label{ProbabPasage}
T_{ij}=\frac{W_{ij}}{\sum_l W_{lj}}.  
\eeq

In this paper we consider biases which are self-consistently related
to graph topological properties.  For instance $W_{ij}$ can be a function
of the vertex properties (the network degree, clustering, etc.) or some
functions of the edge ones (multiplicity or shortest path
betweenness) or any combination of the two. There are other choices of biases
found in the literature such as for instance maximal entropy related
biases~\cite{Bartlomiej}. Some of the results mentioned in this paper hold also for
biases which are not connected to graph properties as will be mentioned in any such case.
Our focus on graph properties for biases is directly connected with application of
biased random walks in examination of community structure in complex networks. 

Let us start by considering a vertex property $x_i$ of the
vertex $i$ (it can be either local as for example the degree, or
related to the first neighbors of $i$ as the clustering coefficient, or global as the
vertex betweenness).  
We choose the following form for the weights: 
\beq\label{ProbBias}
W_{ij}=A_{ij}e^{\beta x_i}\,, \eeq 
where the parameter $\beta\in \mathbb{R}$ tunes the strength 
of the bias. For $\beta=0$ the unbiased case is recovered. By
varying $\beta$ the probability of a walker to move from vertex $j$ 
to vertex $i$ will be enhanced or reduced with respect to the unbiased case
according to the property $x_i$ of the
vertex $i$.  
For instance when $x_i=k_i$, i.e. the degree of the vertex
$i$, for positive values of the parameter $\beta$ the walker will
spend more time on vertices with high degree, i.e. it will
be attracted by hubs.  For $\beta<0$ it will instead
try to ``avoid'' traffic congestion by spending its time on the
vertices with small degree.  The entries of the transition matrix can
now be written as: 
\beq
\label{TransitionEntries}
T_{ij}(\bsy{x},\beta)=\frac{A_{ij}e^{\beta x_i}}{\sum_l A_{lj}e^{\beta
    x_l}}\equiv\frac{A_{ij}e^{\beta x_i}}{z_j(\beta)}.  
\eeq
For this choice of bias we find the following results: (i) we have a
unique representation of any given network via operator
$\bsy{\hat{T}}(\bsy{x},\beta)$, i.e. knowing the operator, we can 
reconstruct the graph; 
(ii) for small
$|\beta|$ we can use perturbation methods around the unbasied case;
(iii) this choice of bias permits in general also to visit vertices with
vanishing feature $x$, which instead is forbidden for instance for a power law 
$W\sim x^\alpha$; (iv) this choice of biases is very common in the studies of energy
landscapes, when biases represent energies $x_i\equiv E_i$ (see for example~\cite{Pollak}
and references therein).

In a similar way one can consider a \emph{symmetric} edge property $y_{ij}$ (for
instance edge multiplicity or shortest path betweenness) as bias. In
this case we can write the transition probability as: 
\beq
\label{TransitionEntriesEdge}
T_{ij}(\bsy{\hat{Y}},\beta)=\frac{A_{ij}e^{\beta y_{ij}}}{\sum_l
  A_{lj}e^{\beta y_{lj}}}.  
\eeq 
The general case of some complicated
multiparameter bias strategy can be finally written as
\beq\label{TransitionEntriesMulti}
T_{ij}(\bsy{x},\bsy{\hat{Y}},\bsy{\beta})=
\frac{A_{ij}e^{\sum_{\nu}\beta_{\nu}x_{i}^{(\nu)}+
    \sum_{\mu}\beta_{\mu}y_{ij}^{(\mu)}}}{\sum_l
  A_{lj}e^{\sum_{\nu}\beta_{\nu}x_{l}^{(\nu)}+\sum_{\mu}\beta_{\mu}y_{lj}^{(\mu)}}}.
\eeq

While we mostly consider biased RW based on vertex properties, as shown below,
most of the results can be extended to the other cases. 
The transition matrix in the former case can also be written as:
$
\bsy{\hat{T}}(x,\beta)=\bsy{\hat{w}}\bsy{\hat{A}}\bsy{\hat{z}}^{-1},
$
where the diagonal matrices $\bsy{\hat{w}}$ and 
$\bsy{\hat{z}}$ are such that 
$w_{ii}=e^{\beta x_i}$ and $z^{-1}_{ii}=1/\sum_l 
A_{li}e^{\beta x_l}$. 
The Frobenius-Perron theorem implies that the largest eigenvalue of
$\bsy{\hat{T}}(x,\beta)$ is always $\l_1(\beta)=1$~\cite{Feller}.  Furthermore, the
eigenvector $\bsy{v}_1$ associated to $\lambda_1$ is strictly positive 
in a connected aperiodic graph.  Its normalized version, denoted as 
$\bsy{p}(\beta)$, gives the asymptotic
stationary distribution of the biased RW on the graph. Assuming for it
the form $p_i(\beta)=\Omega(\beta)^{-1}
g_i(\beta)z_i(\beta)$, where $z_i=\sum_jA_{ij}e^{\beta x_j}$ and
$\Omega(\beta)>0$ is a normalization constant, and plugging this in the
equation $\bsy{p}=\bsy{\hat{T}}\bsy{p}$ we get:
\begin{eqnarray}\label{StatSol}
p_i&=&\sum_jT_{ij}(\bsy{x},\beta)p_j\nonumber\\
&=&\Omega^{-1} e^{\beta x_i}\sum_jA_{ij}g_j\,.
\end{eqnarray}
Hence the equation holds {\it iif} $g_i=e^{\beta x_i}$.  Therefore
the stable asymptotic distribution of vertex centerd biased RWs is
\beq
\label{ExpliDistro} 
p_i(\beta)=\Omega(\beta)^{-1} e^{\beta x_i}z_i(\beta)\,.  \eeq For
$\beta=0$ we have the usual form of the stationary distribution in an
unbiased RW where $z_i(0)=k_i$ and $\Omega(0)=\sum_i k_i$.  For
general $\beta$ it can be easily demonstrated that the asymptotic
solution of edge biased RW is $p_i=\Omega^{-1}z_i$, while for
multiparametric RW the solution is $p_i=\Omega^{-1}
e^{\sum_{\nu}\beta_{\nu}x_i^{(\nu)}}z_i$. 

 Using Eqs. (\ref{ExpliDistro}) and (\ref{TransitionEntries}) we can prove that
the detailed balance condition $T_{ij}p_j=T_{ji}p_{i}$ holds.

At this point it is convenient to introduce a different approach 
to the problem \cite{Blanchard}.
%
%
%
%
%
We start by symmetrizing the matrix $\bsy{\hat{T}}(\bsy{x},\beta)$ 
in the following way:
\begin{equation}\label{SymMat}
\bsy{\hat{T}}^s(\bsy{x},\beta)=[\bsy{\hat{p}}(\beta)]^{-1/2}\bsy{\hat{T}}(\bsy{x
},\beta)
[\bsy{\hat{p}}(\beta)]^{1/2}\,,
\end{equation}
where $\hat{\bf p}(\beta)$ is the diagonal matrix with the stationary
distribution 
$\{p_i(\beta)\}$ on the diagonal.
The entries of the symmetric matrix for vertex centered case are given
by
\begin{equation}\label{SymEnt}
T^{s}_{ij}(\bsy{x},\beta)=
T^{s}_{ji}(\bsy{x},\beta)=A_{ij}\frac{e^{\frac{1}{2}\beta(x_i+x_j)}}{\sqrt{
z_iz_j}}.
\end{equation}
The symmetric matrix 
$\bsy{\hat{T}}^s(\bsy{x},\beta)$ shares the same eigenvalues with the matrix
$\bsy{\hat{T}}(\bsy{x},\beta)$; anyhow the set of eigenvectors
is different and forms a complete orthogonal basis, allowing to define a 
meaningful ``distance`` between vertices. Such distance can provide
important additional information in the problem of community partition of complex
networks.
If ${\bf v}_\nu$ is the $\nu^{th}$ eigenvector of the asymmetric
matrix $\bsy{\hat{T}}(\bsy{x},\beta)$ associated to the eigenvalue 
$\lambda_\nu(\beta)$ (therefore ${\bf v}_1={\bf p}$),
the corresponding eigenvector $\ket{v_\nu}$ of the symmetric matrix
$\bsy{\hat{T}}^s(\bsy{x},\beta)$, can always be written as
$ \ket{v_\nu}_i= \frac{v_{\nu,i}}{\sqrt{p_i}} $.
In particular for $\nu=1$ we have $\ket{v_1}_i\equiv \ket{p}_i=\sqrt{p_i}$.  The same
transformation (\ref{SymMat}) can be applied to the most general multiparametric RW. In
that case the symmetric operator is
\begin{equation}\label{SymEnt2}
T^{s}_{ij}(\bsy{x},\beta)=
T^{s}_{ji}(\bsy{x},\beta)=A_{ij}\frac{e^{(\sum_{\nu}\frac{\beta_{\nu}}{2}(x_{i}^{(\nu)}
+x_ {j}^{(\nu)})+
    \sum_{\mu}\beta_{\mu}y_{ij}^{(\mu)})}}{\sqrt{
z_iz_j}}.
\end{equation}
 
This form also enables usage of perturbation theory for Hermitian linear operators.
For instance, knowing the eigenvalue $\lambda_{\nu}(\beta)$
associated to eigenvector $\ket{v_{\nu}(\beta)}$, we can write the following 
expansions at sufficiently small $\Delta\beta$: $\lambda_\nu(\beta+\Delta\beta)=
\lambda_\nu(\beta)+\Delta\beta\lambda_\nu^{(1)}(\beta)+\ldots$
and $\ket{v_\nu(\beta+\Delta\beta)}=\ket{v_\nu^{(0)}(\beta)}+
\Delta\beta\ket{v_\nu^{(1)}(\beta)}+\ldots$. It follows that for a vertex centered bias
\beq\label{FirstOrderLambda}
\lambda_\nu^{(1)}(\beta)=\bra{v_\nu(\beta)}\bsy{\hat{T}}^{s(1)}(\bsy{x},\beta)
\ket{v_\nu(\beta)},
\eeq 
where, 
\begin{equation}
\label{derivation}
\bsy{\hat{T}}^{s(1)}(\bsy{x},\beta)\!\equiv\!\frac{\partial\bsy{\hat{T}}^s\!(\bsy{
x},\beta)}{
\partial\beta} \!
=\!\frac 1 2\! \left[\lbrace\bsy{\hat{x}},\bsy{\hat{T}}^s
\rbrace_{\small +}\!-\!\lbrace\bsy{\hat{\bar{x}}}(\beta),\bsy{\hat{T}}^s
\rbrace_{\small +}\!\right]
\end{equation}
with $\lbrace\cdot,\cdot\rbrace_{\small +}$ being the anticommutator
operator. Operator $\bsy{\hat{x}}$ and $\bsy{\hat{\bar{x}}}(\beta)$ are
diagonal matrices with $(\bsy{\hat{x}})_{ii}=x_i$ and
$\bsy{\hat{\bar{x}}}_{ii}(\beta)=\sum_l A_{li}x_le^{\beta
  x_l}/z(i)$ which is the expected
value of $x$ that an random walker, will find moving from vertex $i$ to its neighbors.
In the case of edge bias the change of symmetric matrix
with parameter $\beta$ can be written as
$\frac{\partial\bsy{\hat{T}}^s(\beta)}
{\partial\beta}=\bsy{\hat{Y}}\star\bsy{\hat{T}}^s(\beta)-1/2\lbrace\bsy{\hat{\bar{y}}},
\bsy{\hat{T}}^s(\beta)\rbrace_{+}$,
and $\star$ represents the Schur-Hadamard product i.e. element wise multiplication of
matrix elements.
The eigenvector components in $\beta+\Delta\beta$
at the first order of expansion 
in the basis of the eigenvectors at $\beta$ are given by 
(for $\mu \ne \nu$): 
\beq
\label{FirstORdVecs}
\braket{v_\mu(\beta)}{v_\nu^{(1)}(\beta)}=\frac{\bra{v_\mu(\beta)}
  \bsy{\hat{T}}^{s(1)}(\beta)\ket{v_\nu(\beta)}}{\lambda_\mu(\beta)-
  \lambda_\nu(\beta)}.
\eeq
For $\mu=\nu$ the product $\braket{v_\mu(\beta)}{v_\nu^{(1)}(\beta)}$ vanishes
and eqs.~(\ref{derivation}) and (\ref{FirstORdVecs}) 
hold only for non-degenerate cases.
In general, usual quantum mechanical perturbation theory can be used to go 
to higher order perturbations or to take into account
degeneracy of eigenvalues.

We can also exploit further the formal analogy with quantum mechanics using
Parametric Equations of Motion
(PEM)~\cite{Mazziotti95,Mazziotti96} to study the $\beta$
dependence of the spectrum of $\bsy{\hat{T}}^s$.  If we know such spectrum 
for one value of
$\beta$, we can calculate it for any other value of $\beta$ by
solving a set of differential equations corresponding to PEM in
quantum mechanics. They are nothing else the expressions
of Eqs.~(\ref{FirstOrderLambda}) and (\ref{FirstORdVecs}) 
in an arbitrary complete orthonormal base $\{\ket{\phi_\nu}\}$. 
First the eigenvector is expanded in such a
base:
$\ket{v_\nu(\beta)}=\sum \ket{\phi_\xi}
\braket{\phi_\xi}{v_\nu(\beta)}\equiv\sum_\xi c_{\nu
  \xi}(\beta)\ket{\phi_\xi}$.
We can then write
\begin{equation}\label{PEMLambda}
\frac{\partial\lambda_\nu}{\partial\beta}=\bsy{c}_\nu^\intercal(\beta)\frac{
\partial \bsy{\hat{T}}^{s,\phi}(\beta)}
{\partial\beta}\bsy{c}_\nu(\beta)\,, 
\end{equation}
where $\bsy{c}_\nu$ ($\bsy{c}_\nu^\intercal$) is a column (row) vector with 
entries $c_{\nu\xi}(\beta)$ and $\bsy{\hat{T}}^{s,\phi}$ is 
the matrix with entries
$\hat{T}^{s,\phi}_{\nu\xi}(\beta)\equiv\bra{\phi_\nu}\bsy{\hat{T}}
^s(\beta)\ket{\phi_\xi}$.  Let us now define the matrix
$\bsy{\hat{N}}(\beta)$ whose rows are the copies of vector
$\bsy{c}_\nu^\intercal(\beta)$. The differential equation for the
eigenvectors in the basis $\lbrace\ket{\phi_\nu}\rbrace$
is then~\cite{Mazziotti96}
\begin{eqnarray}
\label{PEMvector}
\frac{\partial\bsy{c}_\nu(\beta)}{\partial\beta}\!\!\!&=&\!\left(\bsy{\hat{T}}^{s,
\phi}
(\beta)-\lambda_\nu(\beta)+\bsy{\hat{N}}
(\beta)\right)^{-1}\cdot\nonumber\\
&&\left(\bsy{c}_\nu^\intercal(\beta)\frac{\partial
\bsy{\hat{T}}^{s,\phi}\!(\beta)}{\partial\beta}\bsy{c}_{\nu}(\beta)-
\frac{\partial
\bsy{\hat{T}}^{s,\phi}\!(\beta)}{\partial\beta}\right)\bsy{c}
_\nu(\beta).\nonumber\\
\end{eqnarray}

A practical way to integrate Eqs. (\ref{PEMLambda}) and (\ref{PEMvector}) 
can be found in~\cite{Mazziotti96}. In order to
calculate parameter dependence of eigenvectors and eigenvalues, the best way to
proceed is to perform an LU decomposition of the matrix
$\left(\bsy{\hat{T}}^{s,\phi}(\beta)-
\lambda_\nu(\beta)+\bsy{\hat{N}}(\beta)\right)^{-1}$ as the product of a lower
triangular 
matrix $\bsy{\hat L}$ and an upper triangular matrix $\bsy{\hat U}$, 
and integrate differential equations of higher order which can be
constructed in the same way as equations (\ref{PEMLambda}) and
(\ref{PEMvector})~\cite{Mazziotti96}.
A suitable choice for the basis is just the ordinary unit vectors spanned by
vertices, i.e.
$\lbrace\ket{\phi}\rbrace\equiv\lbrace\ket{e}\rbrace$. We found that for
practical
purposes, depending on the studied network, it is appropriate to use PEM until
the error increases to much and then diagonalize matrix again to get better
precision. PEM efficiently enables study of the large set of parameters for large
networks due to its compatitive advantage over ordinary diagonalization.
\begin{figure}[htbp]
\includegraphics[width=8.0cm,height=6.5cm]{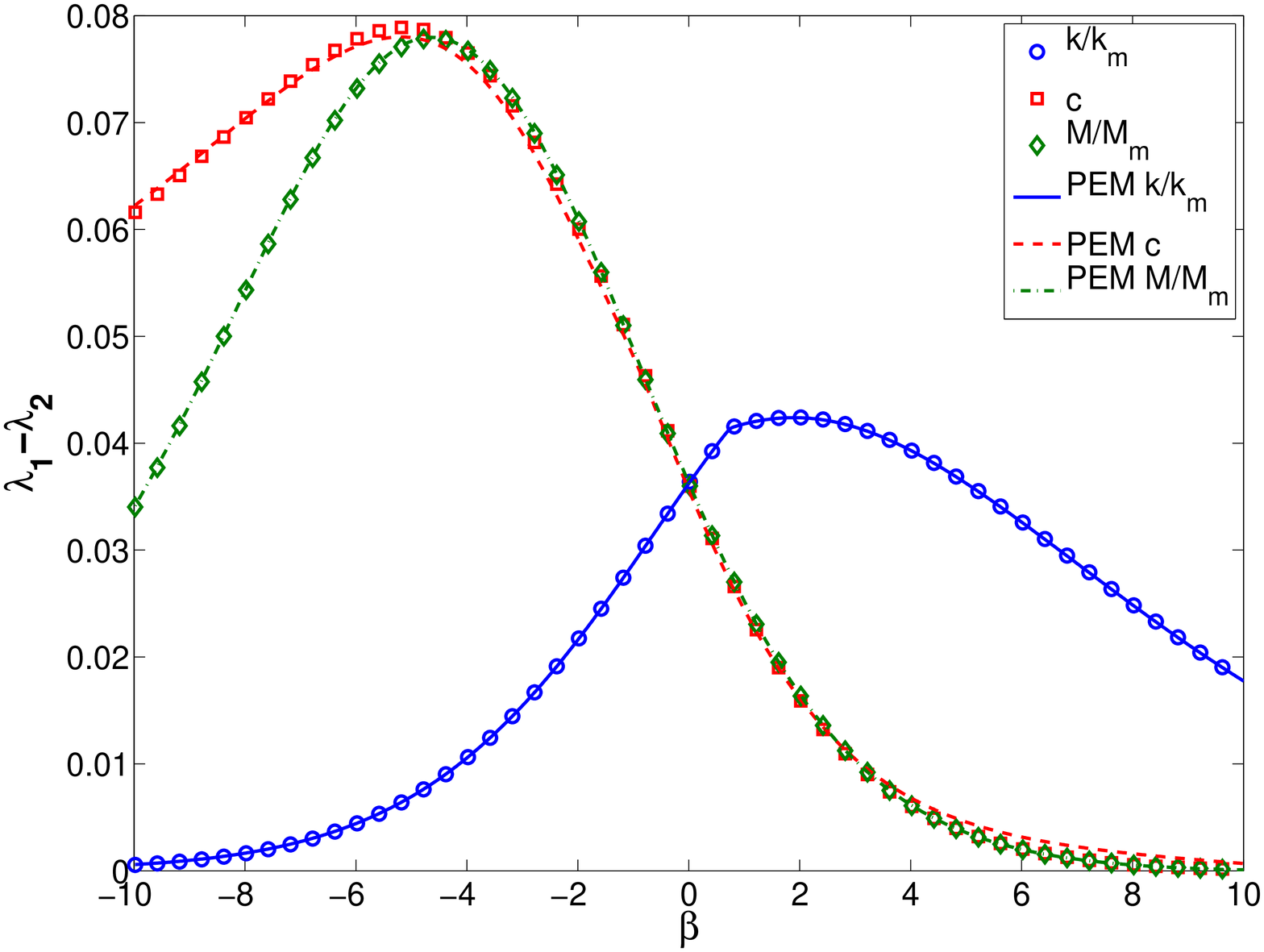}
\caption{
(color on line) Plot of the spectral gap $\lambda_1-
\lambda_2$ vs $\beta$ for
networks of 10 communities with 10 vertices each (the probability for an
edge to be in a community is $p_i=0.3$ while outside of the community it is
$p_o=0.05$). Solid points represent the solutions computed via diagonalization, 
while lines report the value obtained through integration of PEM. Different bias choice 
have been tested. Circles (blue) are related to degree-based strategy, 
square (red) are related to clustering-based strategies, 
diamonds (green) multiplicity-based strategies. 
The physical quantities to get the variable $x$ in Eq.~(\ref{ProbBias})
in these strategies have been 
normalized with respect to their maximum value.}
\label{SpectralGap}
\end{figure}

\section{Spectral Gap}
A key variable in the spectral theory of graphs is the {\it spectral gap}
$\mu=(\lambda_1-\lambda_2)$, i.e. the difference
between first unitary and the second eigenvalues. The spectral gap measures
how fast the information on the RW initial distribution 
is destroyed and the stationary distribution is approached.
The characteristic time for that is
$\tau=-1/\ln[(1-\mu)]\simeq 1/\mu$~\cite{Blanchard}. 
We show in Fig.~1 the dependence of
spectral gap of simulated graphs with communities for different strategies
(degree, clustering and multiplicity based) at a
given value of parameter $\beta$. In all investigated cases the spectral gap has
its well defined maximum, i.e. the value of parameter $\beta$ for which
the random walker converges to stationary distribution with the
largest rate.  

The condition of maximal spectral gap implies that it is a stationary 
point for the function $\lambda_2(\beta)$,
i.e. that its first order perturbation coefficient vanishes at this point:
\begin{eqnarray}\label{maximalSpectralGap}
0&=&\bra{v_2(\beta_m)}\frac{\partial\bsy{\hat{T}}^s(\beta_m)}
{\partial\beta}\ket{v_2(\beta_m)}\nonumber\\ 
&=&\bra{v_2(\beta_m)}\left[\bsy{\hat{x}}-\bsy{\hat{\bar{x}}}
(\beta_m)\right]\ket{v_2(\beta_m)},
\end{eqnarray} 
where $\bsy{\hat{x}}$ and $\bsy{\hat{\bar{x}}}$ are defined above. 
The squares of entries $c_{2,i}^2(\beta)$ of the vector
$\ket{v_2(\beta)}$ in the chosen basis $\ket{\phi_i}\equiv\ket{e}$, 
define a particular measure on
the graph.  Equation (\ref{maximalSpectralGap}) can be written as

$\sum_i c_{2,i}^2(\beta_m)(x_i-\bar{x}_i(\beta_m))=0$. Thus we
conclude that the local spectral maximum is achieved if the average
difference between property $x_i$ and its expectation $\bar{x}_i$, with 
respect to this measure, in the neighborhood of vertex $i$ vanishes. 
We have studied behavior of spectral gap for different sets of real and
simulated networks (Barab\'asi-Albert model with different range of parameters,
Erd\H{o}s-R\'enyi model and random netwroks with given community structure) and three
different strategies (degree-based, 
clustering-based and multiplicity-based). 
Although in general it is not clear that the local
maximum of spectral gap is unique, we have found only one maximum in all the
studied networks. This observation is
interesting because for all cases the shapes of spectral gap {\em vs.} 
$\beta$ looks typically Gaussian-like. In both limits $\beta\to\pm\infty$ the
spectral gap of heterogeneous network is indeed typically zero, as the RW 
stays in the vicinity of the vertices with maximal or minimal value of 
studied property $x_i$. 

\section{Random walk invariant}

A fundamental question in the theory of complex networks is how
topology affects dynamics on networks. Our choice of $\beta$-parametrized 
biases provides a useful tool to investigate 
this relationship. A central issue is, for instance, given by the search of 
properties of the transition matrix $\bsy{T}$
which are independent of $\beta$ and the chosen bias, 
but depend only on the topology of 
the network.  An important example comes from the analysis the determinant 
of $\bsy{T}$ as a function of the bias parameters:
\begin{equation}\label{ProductLambda}
\frac{\partial\prod_{\mu=1}^{N}\lambda_{\mu}}{\partial\beta}=\sum_{\mu=1}^{N}\bra{\mu}
\frac{\partial\bsy{T}^{s}}{\partial\beta}\ket{\mu}\prod_{\nu\neq\mu}\lambda_\nu,
\end{equation}
For vertex centered bias using eq. (\ref{derivation}) we have
\begin{equation}\label{ProdLambda2}
\frac{\partial\prod_{\mu=1}^{N}\lambda_{\mu}}{\partial\beta}=\sum_{\mu=1}^{N}\bra{\mu}
(\bsy{\hat{x}}-\bsy{\hat{\bar{x}}}\ket{\mu}\prod_{\nu=1}^{N}\lambda_\nu.
\end{equation}
and using the diagonality of the $\bsy{\hat{x}}$ and
$\bsy{\hat{\bar{x}}}_{ii}(\beta)=\frac{\partial\ln{z_i}}{\partial \beta}$
\begin{equation}\label{ProdLambda3}
\prod_{\mu=1}^{N}\frac{\lambda_{\mu}(\beta)z_{\mu}(\beta)}{e^{\beta x_{\mu}}}=
\prod_{\mu=1}^{N}\frac{\lambda_{\mu}(\beta_0)z_{\mu}(\beta_0)}{e^{\beta_0 x_{\mu}}}.
\end{equation}
In other words the quantity
$\prod_{\mu=1}^{N}\frac{\lambda_{\mu}(\beta)z_{\mu}(\beta)}{e^{\beta
    x_{\mu}}}$ is a topological constant which does not depend on the
choice of parameters. For $\beta=0$ we get
$\prod_{\mu=1}^{N}\lambda_{\mu}k_\mu=const$ and it follows that this
quantity does not depend on the choice of vertex biases $x_i$
either. It can be shown that such  quantity coincides with the
determinant of adjacency matrix which must be conserved for all
processes. 

\section{Community finding}
\begin{figure}[htbp]
\includegraphics[width=8.0cm,height=6.5cm]{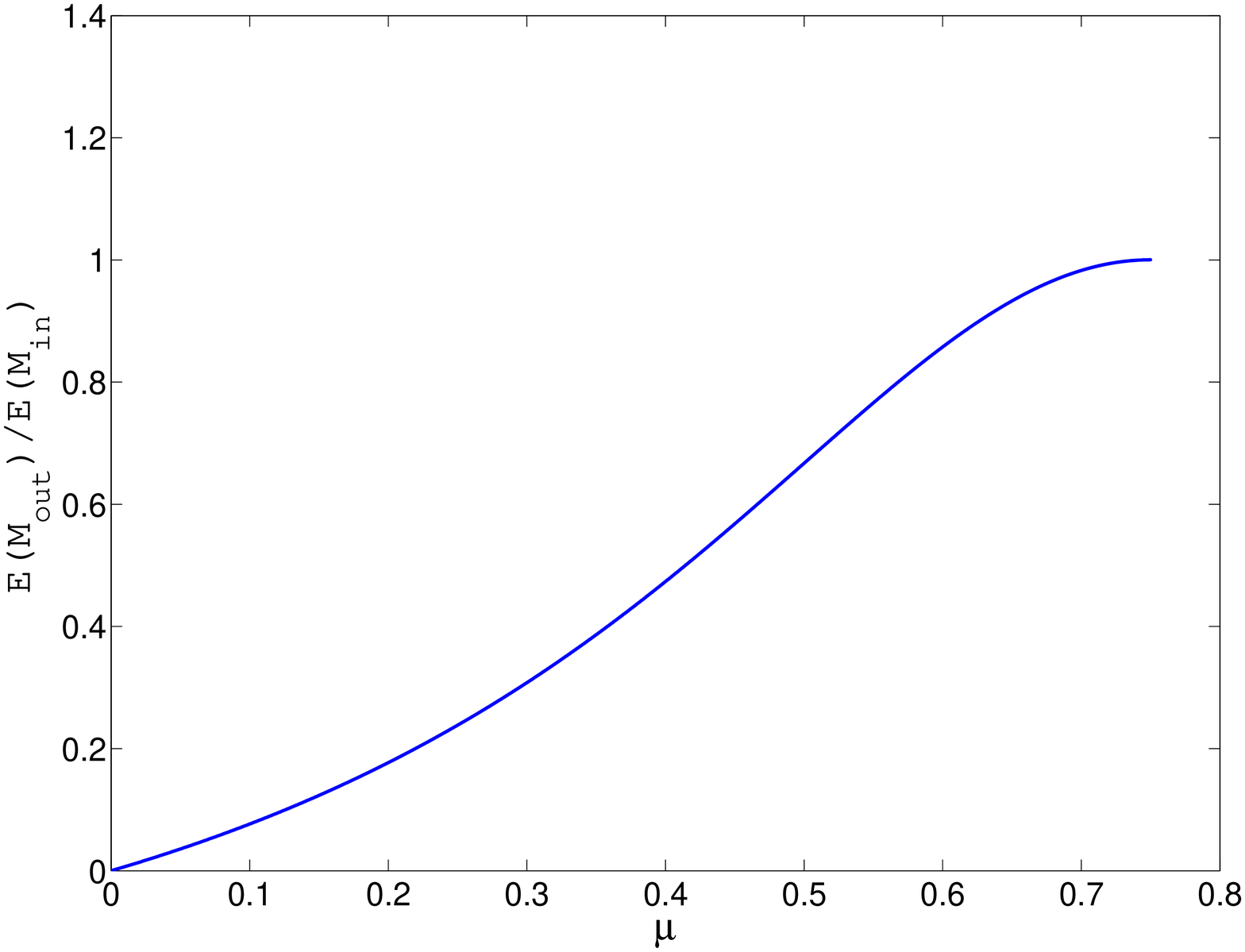}
\caption{
(color on line) The ratio between expected value of multiplicity for edges that are
connecting vertices in different communities to the expected value of multiplicity for
edges that are connecting vertices in the same community with respect to parameter $\mu$.}
\label{params}
\end{figure}

There are many competing algorithms and methods for community
detection~\cite{SantoRev}.  Despite a significant scientific effort to
find such reliable algorithms, there is not yet agreement on a single
general solving algorithm for the various cases. In this section
instead of adding another precise recipe, we want to suggest a general
methodology based on TBRW which could be used for community detection
algorithms.  To add trouble, the very definition of communities is not
a solid one.  In most of the cases we define communities as connected
subgraphs whose density of edges is larger within the proposed
community than outside it (a concept quantified by modularity \cite{ng1} 
).

Scientific community is therefore thriving to find a benchmark in
order to assess the success of various methods. One approach is to
create synthetic graphs with assigned community structure (benchmark
algorithms) and test through them the community detection
recipes \cite{Benchmark}. The Girvan-Newman (GN) \cite{ng1} and
Lancichinetti-Fortunato-Radicchi (LFR) \cite{LFR, Benchmark2} are the most
common benchmark algorithms. In both these models several topological
properties (not only edge density) are unevenly distributed within 
the same community and between different ones.  
We use this property to propose a novel
methodology creating suitable TBRW for community detection.  The
difference between internal and external part of a community is
related to the ``physical'' meaning of the graph. In many real
processes the establishment of a community is facilitated by the
subgraph structure. For instance in social networks agents have a
higher probability of communication when they share a lot of
friends. We test our approach on GN benchmark since in this case we
can easily compute the expected differences between the frequency of
biased variables within and outside the community.

In this section we will describe how to use TBRW for community
detection. For $\beta=0$ our method is rather similar to the one
introduced by Donetti and Mu\~{n}oz \cite{Donetti}. The most notable
difference is that we consider the spectral properties of transition
matrix instead of the Laplacian one. We decide if a vertex belongs to
a community according to the following ideas:  \emph{(i)} We expect
that the vertices belonging to the same community to have similar
values of eigenvectors components; \emph{(ii)} we expect relevant
eigenvectors to have the largest eigenvalues. Indeed, spectral gap is
associated with temporal convergence of random walker fluctuations to
the ergodic stationary state.  If the network has well defined
communities, we expect the random walker to spend some time in the
community rather than escaping immediately out of it. Therefore the
speed of convergence to the ergodic state should be related to the
community structure. Therefore eigenvectors associated with largest
eigenvalues (except for the maximal eigenvalue 1) should be correlated
with community structure. Coming back to the above mentioned Donetti
and Mu\~{n}oz approach here we use the fact that some vertex
properties will be more common inside a community and less frequent
between different communities.  We then vary the bias parameters
trying both to shrink the spectral gap in transition matrix and to
maximize the separation between relevant eigenvalues and the rest of
the spectra.

\begin{figure}[t]
\includegraphics[width=8.0cm,height=6.5cm]{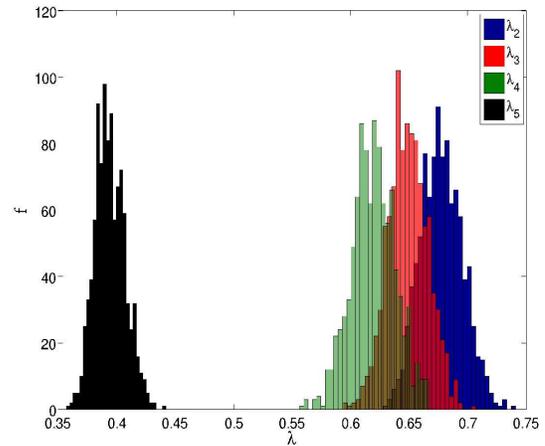}
\caption{
(color on line) Histogram of second, third, fourth and fifth eigenvalue of non-biased
RW for 1000 GN networks with parameters $N=128$, $n=32$, $p_{in}=0.35$, $p_{out}=0.05$.
There is a clear gap between ``community'' band and the rest of the eigenvalues. }
\label{Band}
\end{figure}

\begin{figure}[t]
\includegraphics[width=8.0cm,height=6.5cm]{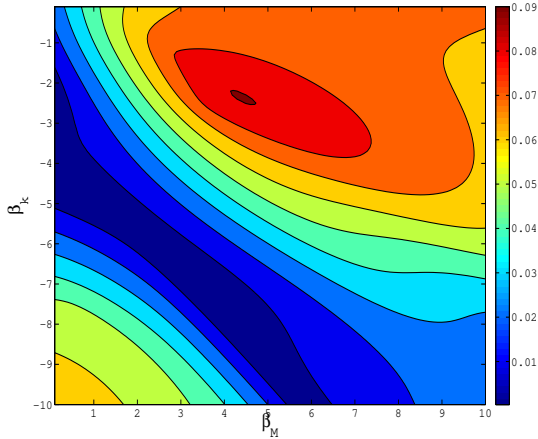}
\caption{
(color on line) Contour plot of the difference between fourth and fifth eigenvalue
$\lambda_4-\lambda_5$ as a function of parameter $\beta_k$ which biases RW 
according to degrees of the vertices and parameter $\beta_M$ which bias RW according to
multiplicities of the edges. Both degrees and multiplicity values are normalized with
respect to the maximal degree and multiplicity (therefore the largest value is one).}
\label{Lambda3Lambda4}
\end{figure}

For example in the case of GN benchmark the network consist of $4$
communities each with $n=32$ vertices i.e. $N=128$ vertices all
together. The probability that the two vertices which belong to the
same community are connected is $p_{in}$. The probability that the two
vertices which belong to different communities are connected is
$p_{out}$. The fundamental parameter \cite{SantoRev} which characterizes
the difficulty of detecting the structure is \beq\label{MU}
\mu=\frac{\bar{k}_{out}}{\bar{k}_{out}+\bar{k}_{in}}, \eeq where
$\bar{k}_{out}=p_{out}(N-n)$ is the mean degree related to
inter-community connections and $\bar{k}_{in}=p_{in}(n-1)$ is the mean
degree related to edges inside-community.  As a rule of thumb we can
expect to find well defined communities when $\mu<1/2$, and observe some
signature of communities even when $\mu<3/4$~\cite{Benchmark}.  The
probabilities $p_{in}$ and $p_{out}$ are related via the control parameter
$\mu$ as $p_{out}=\frac{(n-1)\mu}{(N-n)(1-\mu)}p_{in}$.

We now examine the edge multiplicity. The latter is defined as the
number of common neighbors shared by neighbouring vertices.  The
expected multiplicity of an edge connecting vertices inter-community
and inside-communities are respectively
\begin{eqnarray}\label{expMultIn}
E(M_{out})&=&2p_{in}p_{out}(n-1)+p_{out}^2(N-2n),\nonumber\\
E(M_{in})&=&p_{in}^2(n-2)+p_{out}^2(N-n)
\end{eqnarray}
On Fig.~\ref{params} we plot the ratio of the quantitites above defined, 
$E(M_{out})/E(M_{in})$, {\it vs.} the parameter $\mu$.

We see that even for $\mu>0.5$ the ratio remains smaller than $1$
implying that the multiplicity is more common in the edges in the same
community. Based on this analysis for this particular example we
expect that if we want to find well-defined communities via TBRW we
have to increase bias with respect to the multiplicity. Through
numerical simulations we find that the number of communities is
related to number of eigenvalues in the ``community band''. Namely one
in general observes a gap between eigenvalues
$\lambda_2,...,\lambda_{N/n-1}$ and the next eigenvalue evident in a
network with a strong community structure ($\mu\ll 1/2$).  The
explanation that we give for that phenomenon can be expressed by
considering a network of $n$ separated graphs.  For such a network
there are $n$ degenerate eignevalues $\lambda_1\ldots\lambda_n=1$. If
we now start to connect these graphs with very few edges such a
degeneracy is broken with the largest eigenvalue remaining $1$ while
the next $(n-1)$ eigenvalues staying close to it. The distance between
any two of this set of $(n-1)$ eigenvalues will be smaller than the
gap between this community band and the rest of the eigenvalues in the
spectrum.  Therefore, the number of eigenvalues different of $1$ which
are forming this ``community band'' is always equal to the number of
communities minus one, at least for different GN-type networks with
different number of communities and different sizes, as long as
$\mu\ll 1/2$.  For example in the case of 1000 GN networks described
with parameters $N=128$, $n=32$, $p_{in}=0.35$, $p_{out}=0.05$,
i.e. $\mu=0.125$, the histograms of eigenvalues are depicted on figure
\ref{Band}.

For our purposes we used two parameters biased RW, in which
topological properties are $x_i\equiv k_i/max(k)$ i.e. the normalized
degree (with respect to maximal degree in the network) and
$y_{ij}=M_{ij}/max(M_{ij})$ i.e. the normalized multiplicity (with
respect to maximal multiplicity in the network). We choosed GN network
whose parameters are $N=128$, $n=32$, $p_{in}=16/62$ and
$p_{out}=1/12$, for which $\mu=1/2$.  Being $N/n=4$ the number of
communities, as a criterion for good choice of parameters we decided
to use the difference between $\lambda_4$ and $\lambda_5$, i.e., we
decided to maximize the gap between ``community band'' and the
rest of eigenvalues; checking at the same time that the spectral
gap shrinks.  In Fig.~\ref{Lambda3Lambda4}, we plot such a quantity with
respect to different biases.

\begin{figure}[t]
\includegraphics[width=8.0cm,height=6.5cm]{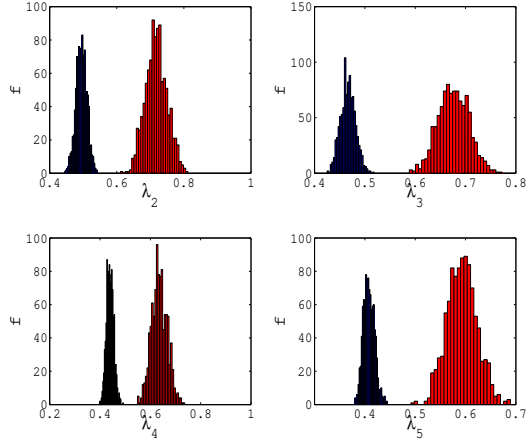}
\caption{ (color on line) Histograms of $\lambda_2$, $\lambda_3$,
  $\lambda_4$ and $\lambda_5$ for 1000 GN networks described with
  parameters $N=128$, $n=32$, $p_{in}=16/62$ and $p_{out}=1/12$. With
  black colour we indicate the eigenvalues of nonbiased RW, while with
  red we indicate the eigenvalues of RW biased with parameters
  $\beta_k=-2.5$ and $\beta_M=4.3$. Note how this choice of parameters
  does not maximize ``community gap'' for all the different
  realizations of monitored GN network.}
\label{LambdaHistograms}
\end{figure}

\begin{figure}[t]
\includegraphics[width=8.0cm,height=6.5cm]{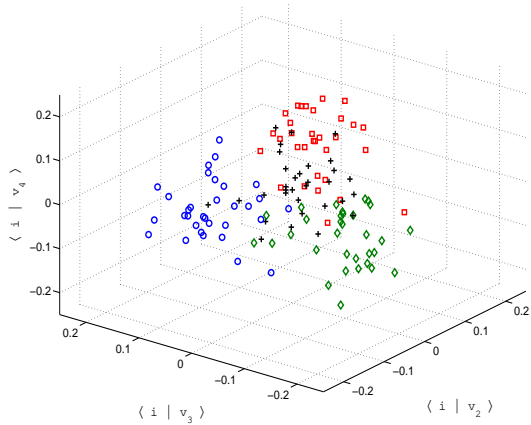}
\caption{ (color on line) Plot of the eigenvector components of the
  second, third and fourth eigenvector. Different markers represent four different
predefined communities. This is an example of GN graph
  with $p_{in}=16/62$ and $p_{out}=1/12$. For this choice of
  parameters $\mu=1/2$. There is a strong dispersion between different
  vertices which belong to the same community.}
\label{Unbiased}
\end{figure}

\begin{figure}[t]
\includegraphics[width=8.0cm,height=6.5cm]{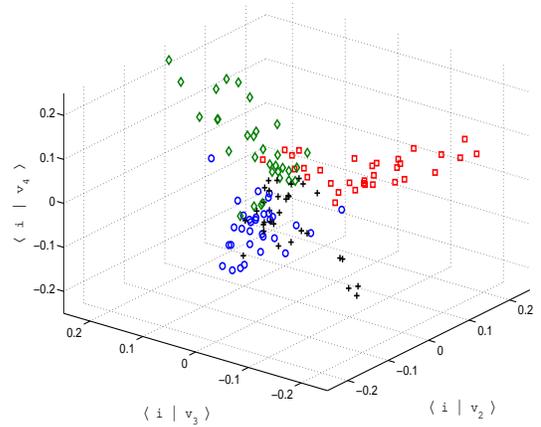}
\caption{(color on line) Plot of the eigenvector components of the
  second, third and fourth eigenvector of biased RW with parameters
  $\beta_k=-2.5$ and $\beta_M=4.3$. Different markers represent four different
predefined communities. This is an example of the same GN graph realization
  with $p_{in}=16/62$ and $p_{out}=1/12$ as the
  one on the previous figure. For this choice of parameters
  $\mu=1/2$. One can notice tetrahedral distribution of vertices in
  which vertices from the same community belong to the same branch of
  tetrahedron.}
\label{Biased}
\end{figure}

It is important to mention that for every single network instance
there are different optimal parameters.  This can be seen on figure
\ref{LambdaHistograms}, where we show the difference between
unbiased and biased eigenvalues for 1000 GN nets created with same
parameters.  As shown in the figure the difference between fourth and
fifth eigenvalue is now not necessarily the optimal for this choice of
parameters. Every realization of the network should be independently
analyzed and its own parameters should be carefully chosen.
In the Figs.~\ref{Unbiased} and \ref{Biased} we present instead the
difference between unbiased and biased projection on three
eigenvectors with largest nontrivial eigenvalues. Using 3D view it is
easy to check that communities are better separated in the biased case
then in the non-biased case.

\section{Conclusion}

In this paper we presented a detailed theoretical framework to analyze
the evolution of TBRW on a graph. By using as bias some topological
property of the graph itself allows to use the RW as a tool to explore
the environment. This method maps vertices of the graph to different
points in the $N$-dimensional Euclidean space naturally associated
with the given graph. In this way we can measure distances between
vertices depending on the chosen bias strategy and bias parameters. In
particular we developed a perturbative approach to the spectrum of
eigenvalues and eigenvectors associated with the transition matrix of
the system. More generally we generalized the quantum PEM approach to
the present case. This led naturally to study the behavior of the gap
between the largest and the second eigenvalue of the spectrum
characterizing the relaxation to the stationary Markov state.  In
numerical applications of such a theoretical framework we have
observed a unimodal shape of the spectral gap {\em vs.} the bias
parameter which is not an obvious feature of the studied processes.
We have finally outlined a very promising application of topologically
biased random walks to the fundamental problem of community finding.
We described the basic ideas and proposed some criteria for the choice of
parameters, by considering the particular case of GN graphs. 
We are working further in direction of this application,
but the number of possible strategies (different topological
properties we can use for biasing) and types of networks is just too
large to be presented in one paper.  Furthermore, since in many
dynamical systems as the WWW or biological networks, a feedback
between function and form (topology) is evident, our framework may be
a useful way to describe mathematically such an observed mechanism.
In the case of biology, for instance, the shape of the metabolic
networks can be triggered not only by the chemical properties of the
compounds, but also by the possibility of the metabolites to
interact. Biased RW can be therefore the mechanism through which a
network attains a particular form for a given function. By introducing
such approach we can now address the problem of community detection in
the graph. This the reason why here we have not introduced another precise 
method for community detection, but rather a possible framework 
to create different community finding methods with different {\em ad hoc} 
strategies. Indeed in real situations we expect
different types of network to be efficiently explored by use of
different topological properties. This explains why we believe that TBRW
could play a role in community detection problems, and we hope to
stimulate further developments, in the network scientific community, of this
promising methodology.

{\bf Acknowledgments}
Vinko Zlati\'c wants to thanks MSES of the Republic of Croatia through 
project No. 098-0352828-2836 for partial support. Authors acknowledge support from EC FET
Open Project "FOC" nr. 255987.

\end{document}